\input{aipcheck}

\documentclass[
    ,final            
  ]
  {aipproc}

\layoutstyle{6x9}

\begin{document}

\title{Browsing the sky through the ASI Science Data Centre Data Explorer Tool}


\author{V. D'Elia, M. Capalbi, F. Verrecchia, B. Gendre, P. Giommi}{
 address={ASI Science Data Centre, Via Galileo Galilei, 00044 Frascati (RM) I}
}

\begin{abstract}
  We present here the Data Explorer tool developed at the ASI Science
  Data Center (ASDC). This tool is designed to provide an efficient
  and user-friendly way to display information residing in several
  catalogs stored in the ASDC servers, to cross-correlate this
  information and to download/analyze data via our scientific tools
  and/or external services. Our database includes GRB catalogs (such
  as Swift and Beppo-SAX), which can be queried through the Data
  Explorer. The GRB fields can be viewed in multiwavelength and the
  data can be analyzed or retrieved.

\end{abstract}

\maketitle


\section{Introduction}

Data Explorer is a tool which allows the user to navigate through the
ASDC and external catalogs and provides an easy way to access data. It
can be accessed through the ASDC home page (www.asdc.asi.it), from the
'Quick Look Data', 'Multi mission Archive' and 'Tools' main menu tabs,
and from the 'Archive Explorer' panel.

\section{Getting Started}

If you reach the Data Explorer, you have made a query to the ASDC
system for a specific source or set of coordinates. The source
position in the sky is displayed on the top right corner of
fig. 1. The input coordinates are shown at the top of the page and
below the graphic frame, enclosed in a yellow, red-bordered
rectangle. These coordinates are given both in RA/Dec (decimal and
sexagesimal) and l/b systems, and represent the center of a 60X60
arcmin box opened by default.

This graphical box displays all the sources collected in the ASDC
default catalogs. These sources can be listed clicking the 'Source
list' link below the grafical box. Information about each source can
also be gathered moving the mouse on the circle corresponding to that
specific source.The default catalogs can be viewed clicking the 'i'
tab close to 'Default catalogs'.

\begin{figure}
  \includegraphics[height=.5\textheight]{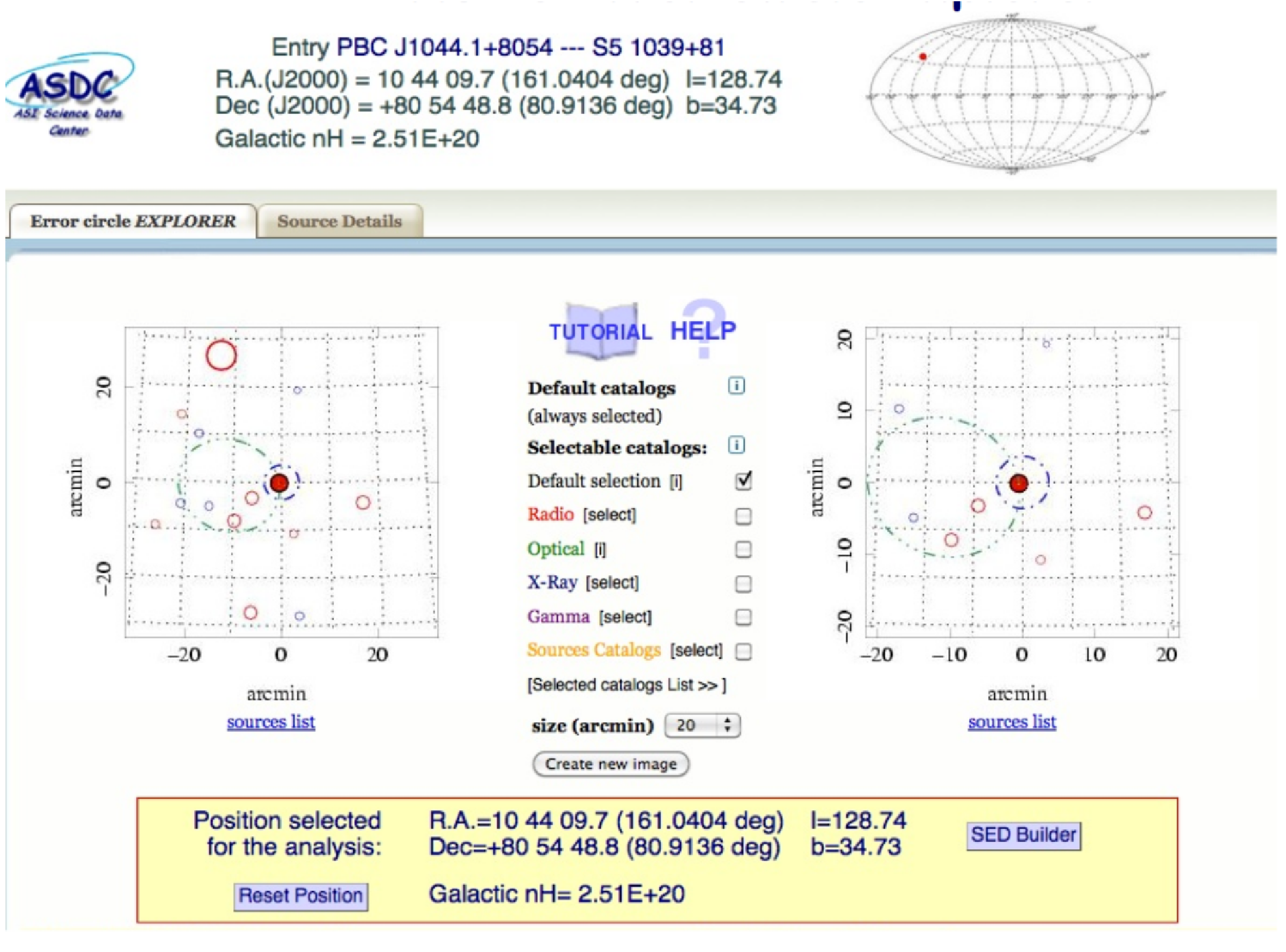}
  \caption{The ASDC Error Circle Explorer}
\end{figure}

\section{Refining the search criteria}

The database search can be refined. Queried catalogs can be selected
using the central panel in the figure below. ASDC selectable catalogs
are listed by band (from lower to higher energies) and can be selected
clicking the corresponding link. The list of chosen catalogs can be
retrieved using the 'Selected catalogs List' button. A full list of
all selectable catalogs is also supplied.

In addition, clicking on a specific source on the left panel, will
cause the coordinates in the lower box to change. The 'Create new
image' button will generate a new plot (right panel) centered on the
new position. The spatial coverage of the right plot can be adjusted using
the 'size (arcmin)' toolbox. Further centerings can now be performed
also clicking on the right plot.  Source colors refer to catalogs
according to their wavelengths, coherently with the central panel
colors. The red cross in the right box represents the original set of
coordinates. Finally, the new query can also be executed on a new
graphical window, and a spectral energy distribution for the selected
source can be accessed by clicking the 'SED Builder' tab, located in
the right region of the rectangular yellow box.  The SED tool tutorial
can be found at:
http://tools.asdc.asi.it/SED/docs/SEDTool-Tutorial.pdf

\section{Additional services}

The 'Search ASDC Catalogs' panel (fig. 2, top) allows the user to
browse the internal catalogs (grouped by energy band) for sources
around the current coordinates, in a user-defined search radius.  The
'Search Other Services' panel queries external databases using the
current coordinates and in a user-defined search radius. Both services
display the results in a new window.

\begin{figure}
  \includegraphics[height=.17\textheight]{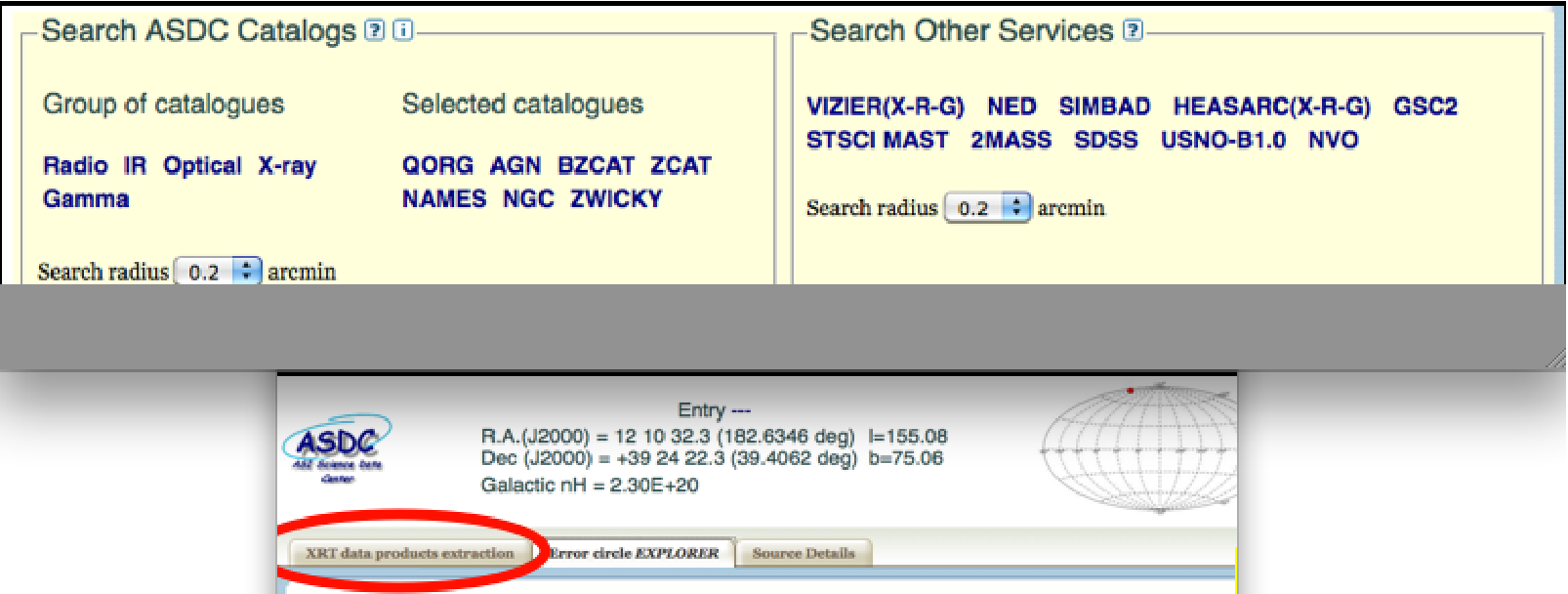}
  \includegraphics[height=.29\textheight]{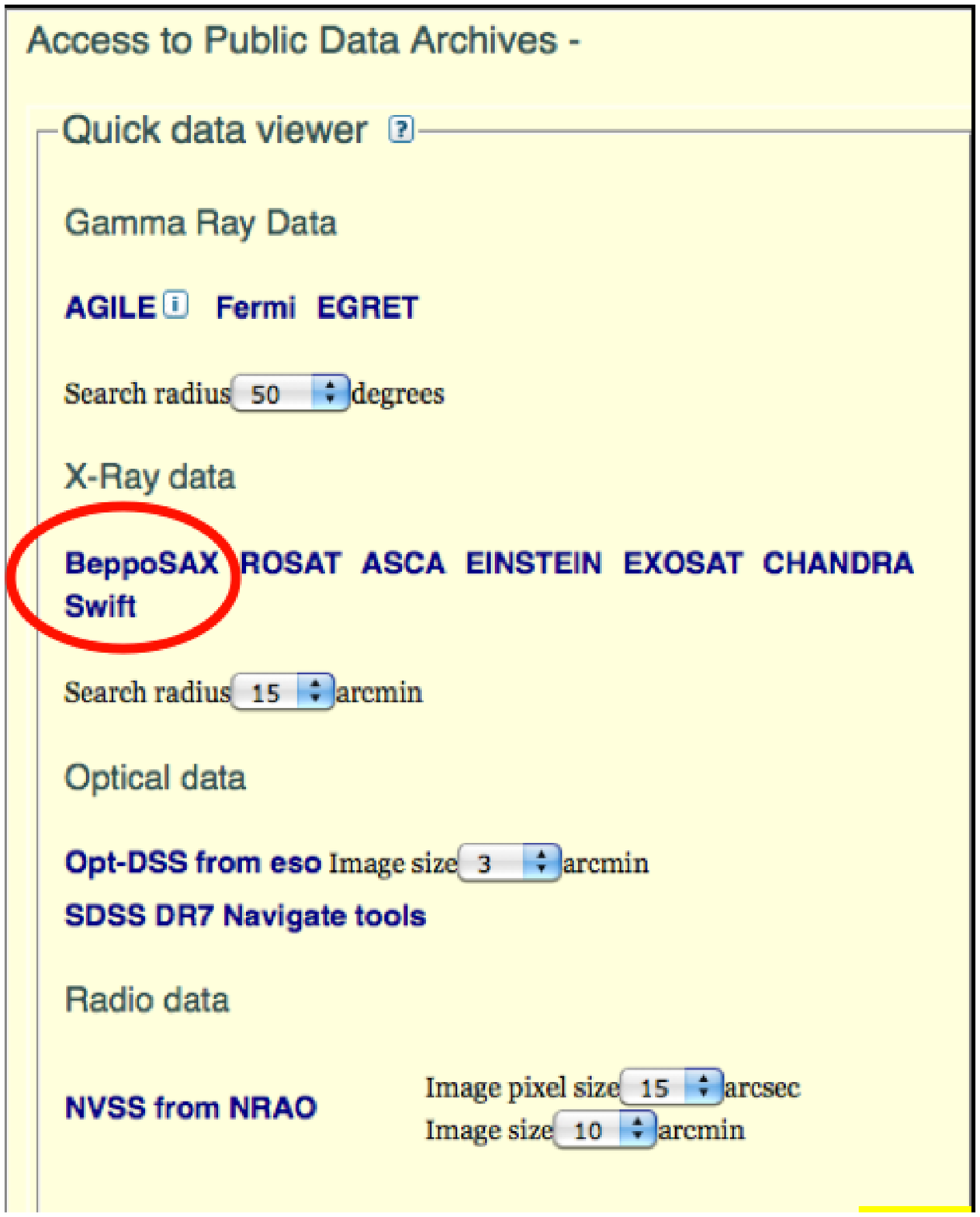}
  \caption{Additional services and ccess to public data}
\end{figure}

\section{Access to public data archives}

The quick data viewer (fig. 2, right) allows the user to search and
download data from specific missions/databases in a user-defined
search radius around the current coordinates.  Missions and databases
are ordered from top to bottom as the energy band decreases.

\section{Data Explorer and GRBs}

In the GRB framework Data Explorer can be used to: i) quickly check
the coordinates of GRB triggers to verify the presence of possible
counterparts at all wavelength bands and thus recognize triggers from
known, variable sources; ii) examine the GRB fields to search for host
galaxies, host candidates or possible intervening absorbers on the
line of sight; iii) access the BeppoSAX and Swift archives to retrieve
data and/or perform (red ellipses in fig. 2) a quick online analysis
(see Stratta et al. 2010, these proceedings for a tutorial of the
Swift/XRT online analysis that can be performed at the ASDC website).

\section{Credits}

Data Explorer is a tool provided by the ASI Science Data Centre. For
any bug or request please refer to:
http://www.asdc.asi.it/feedback\_all. If you need assistance:
http://swift.asdc.asi.it/helpdesk/login.php?cat=generic. The full
documentation can be found at:
http://www.asdc.asi.it/tutorial/DataExplorer/DataExplorerTutorial.html





\bibliographystyle{aipproc}   

\bibliography{sample}

\IfFileExists{\jobname.bbl}{}
 {\typeout{}
  \typeout{******************************************}
  \typeout{** Please run "bibtex \jobname" to optain}
  \typeout{** the bibliography and then re-run LaTeX}
  \typeout{** twice to fix the references!}
  \typeout{******************************************}
  \typeout{}
 }

\end{document}